\documentclass[conference]{IEEEtran}
\IEEEoverridecommandlockouts

\usepackage{cite}
\usepackage{amsmath,amssymb,amsfonts}
\usepackage{algorithmic}
\usepackage{graphicx}
\usepackage{textcomp}
\def\BibTeX{{\rm B\kern-.05em{\sc i\kern-.025em b}\kern-.08em
    T\kern-.1667em\lower.7ex\hbox{E}\kern-.125emX}}
    
\DeclareGraphicsExtensions{.pdf,.jpeg,.png,.svg}
\usepackage[table,xcdraw]{xcolor}
\usepackage[caption=false,font=footnotesize]{subfig}
\usepackage{tabularx}
\usepackage[hidelinks]{hyperref}
\usepackage{booktabs}
\usepackage{orcidlink}

\newcommand\copyrighttext{%
  \footnotesize \textcopyright 2022 IEEE. Personal use of this material is permitted. Permission from IEEE must be obtained for all other uses, in any current or future media, including reprinting/republishing this material for advertising or promotional purposes, creating new collective works, for resale or redistribution to servers or lists, or reuse of any copyrighted component of this work in other works. IEEE Copyright policy can be found at \textit{\href{https://www.ieee.org/publications/rights/copyright-policy.html}{https://www.ieee.org/publications/rights/copyright-policy.html}}
  }
\newcommand\copyrightnotice{%
\begin{tikzpicture}[remember picture,overlay]
\node[anchor=south,yshift=10pt] at (current page.south) {\fbox{\parbox{\dimexpr\textwidth-\fboxsep-\fboxrule\relax}{\copyrighttext}}};
\end{tikzpicture}%
}

\begin{document}

%%%%%%%%%%%%%%%%%%%%%%%%%%%%%%%%%%%%%%%%%%%%%%%%%%%%%%%%%%%%%%%%%%%%%%%%%%%%%%%%%%%%%%%
%%%%%%%%%%%%%%%%%%%%%%%%%%%%%%%%%%%%% Title %%%%%%%%%%%%%%%%%%%%%%%%%%%%%%%%%%%%%%%%%%%
%%%%%%%%%%%%%%%%%%%%%%%%%%%%%%%%%%%%%%%%%%%%%%%%%%%%%%%%%%%%%%%%%%%%%%%%%%%%%%%%%%%%%%% 
\title{Virtual Sensor Middleware: Managing {IoT} Data for the Fog-Cloud Platform
\thanks{\hrule\vspace{2pt}\noindent This research has been supported by NSERC under grants RPGIN-2017-02461 \& RGPIN-2018-06222.}
}
%%%%%%%%%%%%%%%%%%%%%%%%%%%%%%%%%%%%%%%%%%%%%%%%%%%%%%%%%%%%%%%%%%%%%%%%%%%%%%%%%%%%%%%
%%%%%%%%%%%%%%%%%%%%%%%%%%%%%%%%%%%%%%%%%%%%%%%%%%%%%%%%%%%%%%%%%%%%%%%%%%%%%%%%%%%%%%% 

%%%%%%%%%%%%%%%%%%%%%%%%%%%%%%%%%%%%%%%%%%%%%%%%%%%%%%%%%%%%%%%%%%%%%%%%%%%%%%%%%%%%%%%
%%%%%%%%%%%%%%%%%%%%%%%%%%%%%%%%%%%%% Authors %%%%%%%%%%%%%%%%%%%%%%%%%%%%%%%%%%%%%%%%%
%%%%%%%%%%%%%%%%%%%%%%%%%%%%%%%%%%%%%%%%%%%%%%%%%%%%%%%%%%%%%%%%%%%%%%%%%%%%%%%%%%%%%%% 
\author{
\IEEEauthorblockN
{
    % Fadi AlMahamid \orcidlink{0000-0002-6907-7626}, Hanan Lutfiyya \orcidlink{0000-0002-5341-9388}, Katarina Grolinger \orcidlink{0000-0003-0062-8212}
    Fadi AlMahamid, Hanan Lutfiyya, Katarina Grolinger
}
\IEEEauthorblockA
{
    \textit{Western University}\\
    London, Ontario, Canada\\
    Email: \{falmaham, hlutfiyy, kgroling\}@uwo.ca\\
    ORCID: \orcidlink{0000-0002-6907-7626} 0000-0002-6907-7626, \orcidlink{0000-0002-5341-9388} 0000-0002-5341-9388, \orcidlink{0000-0003-0062-8212} 0000-0003-0062-8212
}
}

%%%%%%%%%%%%%%%%%%%%%%%%%%%%%%%%%%%%%%%%%%%%%%%%%%%%%%%%%%%%%%%%%%%%%%%%%%%%%%%%%%%%%%%
%%%%%%%%%%%%%%%%%%%%%%%%%%%%%%%%%%%%%%%%%%%%%%%%%%%%%%%%%%%%%%%%%%%%%%%%%%%%%%%%%%%%%%%

\maketitle
\copyrightnotice

%%%%%%%%%%%%%%%%%%%%%%%%%%%%%%%%%%%%%%%%%%%%%%%%%%%%%%%%%%%%%%%%%%%%%%%%%%%%%%%%%%%%%%%
%%%%%%%%%%%%%%%%%%%%%%%%%%%%%%%%%%%% Abstract %%%%%%%%%%%%%%%%%%%%%%%%%%%%%%%%%%%%%%%%%
%%%%%%%%%%%%%%%%%%%%%%%%%%%%%%%%%%%%%%%%%%%%%%%%%%%%%%%%%%%%%%%%%%%%%%%%%%%%%%%%%%%%%%% 
\vspace{-10pt}
\begin{abstract}
This paper introduces the Virtual Sensor Middleware (VSM), which facilitates distributed sensor data processing on multiple fog nodes. VSM uses a Virtual Sensor as the core component of the middleware. The virtual sensor concept is redesigned to support functionality beyond sensor/device virtualization, such as deploying a set of virtual sensors to represent an IoT application and distributed sensor data processing across multiple fog nodes. Furthermore,  the virtual sensor deals with the heterogeneous nature of IoT devices and the various communication protocols using different adapters to communicate with the IoT devices and the underlying protocol. VSM uses the publish-subscribe design pattern to allow virtual sensors to receive data from other virtual sensors for seamless sensor data consumption without tight integration among virtual sensors, which reduces application development efforts. Furthermore, VSM enhances the design of virtual sensors with additional components that support sharing of data in dynamic environments where data receivers may change over time, data aggregation is required, and dealing with missing data is essential for the applications.
\end{abstract}
%%%%%%%%%%%%%%%%%%%%%%%%%%%%%%%%%%%%%%%%%%%%%%%%%%%%%%%%%%%%%%%%%%%%%%%%%%%%%%%%%%%%%%%
%%%%%%%%%%%%%%%%%%%%%%%%%%%%%%%%%%%%%%%%%%%%%%%%%%%%%%%%%%%%%%%%%%%%%%%%%%%%%%%%%%%%%%%

%%%%%%%%%%%%%%%%%%%%%%%%%%%%%%%%%%%%%%%%%%%%%%%%%%%%%%%%%%%%%%%%%%%%%%%%%%%%%%%%%%%%%%%
%%%%%%%%%%%%%%%%%%%%%%%%%%%%%%%%%%%% Keywords %%%%%%%%%%%%%%%%%%%%%%%%%%%%%%%%%%%%%%%%%
%%%%%%%%%%%%%%%%%%%%%%%%%%%%%%%%%%%%%%%%%%%%%%%%%%%%%%%%%%%%%%%%%%%%%%%%%%%%%%%%%%%%%%%
%%%%%%%%%%%%%%%%%%%%%%%%%%%%%%%%%%%%%%%%%%%%%%%%%%%%%%%%%%%%%%%%%%%%%%%%%%%%%%%%%%%%%%%
%%%%%%%%%%%%%%%%%%%%%%%%%%%%%%%%%%%%%%%%%%%%%%%%%%%%%%%%%%%%%%%%%%%%%%%%%%%%%%%%%%%%%%%

%%%%%%%%%%%%%%%%%%%%%%%%%%%%%%%%%%%%%%%%%%%%%%%%%%%%%%%%%%%%%%%%%%%%%%%%%%%%%%%%%%%%%%%
%%%%%%%%%%%%%%%%%%%%%%%%%%%%%%%%%%%%%%%%%%%%%%%%%%%%%%%%%%%%%%%%%%%%%%%%%%%%%%%%%%%%%%% 
\section{Introduction} \label{sec:intro}
%%%%%%%%%%%%%%%%%%%%%%%%%%%%%%%%%%%%%%%%%%%%%%%%%%%%%%%%%%%%%%%%%%%%%%%%%%%%%%%%%%%%%%%
%%%%%%%%%%%%%%%%%%%%%%%%%%%%%%%%%%%%%%%%%%%%%%%%%%%%%%%%%%%%%%%%%%%%%%%%%%%%%%%%%%%%%%%
The Internet of Things (IoT) is a concept that depicts the world as a realm where physical objects (things) equipped with sensors, actuators, and network connections may communicate and respond to their environment \cite{alarbi2018sensing}. These “things” typically provide data, act on the environment, and encompass control points. Many of these “things” will be able to generate data continuously. As the cost of sensors and local/wireless network connectivity becomes less expensive, there is an increased interest in applications. Sensor data is seen as providing opportunities for new services, improved efficiency, and possibly more competitive business models in a variety of application domains, e.g., smart cities and smart transportation \cite{zanella2014internet}.

Most Internet-based applications access remote computing resources through cloud data centers \cite{barcelo2016iot}. However, this computing model is unsuitable for IoT applications requiring large amounts of geographically dispersed data and latency-sensitive applications \cite{shahid2018iot}. \textit{Fog computing} is a computing model that augments cloud resources with computing and storage resources placed closer to the network edge, which enables some of the processing of computing nodes to be closer to the data sources. Therefore, it reduces the amount of data sent to the cloud and enables faster processing since data does not have to be sent to the cloud. Fog nodes are distributed, heterogeneous, and have resource-constrained computing nodes compared to cloud computing resources. To best leverage fog computing resources, IoT applications often use a {\it dataflow model} that conceptualizes an IoT application as a graph where nodes represent services and edges represent a flow between services. 

With the dataflow model, multiple applications could use the same instance of a service that collects data from a physical sensor or a service that is a component in a pipeline of services starting from the services directly reading from the physical sensors. For example, surveillance video feeds may be used to locate a missing child or crowd count. Image analysis requires feature extraction, which includes various image preprocessing techniques (e.g., binarization, thresholding, resizing, normalization) applied to a  sampled image. Then, the extracted features are used in classifying and recognizing images that are helpful in various image processing applications. The image preprocessing and feature extraction techniques would be typical for the missing child or crowd count. The services should be able to send data to other multiple services and accommodate changes in services to receive the data. A service may require data from multiple services that might arrive at different rates, where some of the data may be missing or corrupted.

The dataflow model often assumes using a {\it virtual sensor} \cite{Aberer2006}. Most of the work on virtual sensor design focuses on receiving data from multiple physical sensors and applying a function to the received data to produce a new measurement. This paper aims to enhance the design of virtual sensors with additional components that support sharing of data in dynamic environments where data receivers may change over time, data aggregation is required, and dealing with missing data is essential for the applications. The enhanced virtual sensor can be used to receive data from a physical sensor or another service.

% This paper presents the \textit{Virtual Sensor Middleware (VSM)} that deploys virtual sensors representing IoT application services based on an innovative design of a virtual sensor.  
This paper presents the \textit{Virtual Sensor Middleware (VSM)} that deploys virtual sensors representing IoT application services based on an innovative design of a virtual sensor. VSM has the following contributions: 1) Decouples IoT applications and services from the physical sensors by seamlessly subscribing to pre-processed data 2) Improves data latency by distributing the processing of IoT data on edge nodes 3) Generates a personalized context-aware sensor data 4) Provides highly configurable middleware that can serve various requirements by diverse IoT applications.

The rest of the paper is organized as follows: Section \ref{sec:RelatedWork} describes the related work. Section \ref{sec:arch} discuss the middleware architecture. Section \ref{sec:interactions} describes components interaction with each other. Section \ref{sec:Implementation} explains the middleware implementation technology. Section \ref{sec:Evaluation} describes the experiments setup and the evaluation results. Finally, Section \ref{sec:Conclusion} concludes the paper and highlights future work.

%%%%%%%%%%%%%%%%%%%%%%%%%%%%%%%%%%%%%%%%%%%%%%%%%%%%%%%%%%%%%%%%%%%%%%%%%%%%%%%%%%%%%%%
%%%%%%%%%%%%%%%%%%%%%%%%%%%%%%%%%%%%%%%%%%%%%%%%%%%%%%%%%%%%%%%%%%%%%%%%%%%%%%%%%%%%%%%  
\section{Related Work} \label{sec:RelatedWork}
%%%%%%%%%%%%%%%%%%%%%%%%%%%%%%%%%%%%%%%%%%%%%%%%%%%%%%%%%%%%%%%%%%%%%%%%%%%%%%%%%%%%%%%
%%%%%%%%%%%%%%%%%%%%%%%%%%%%%%%%%%%%%%%%%%%%%%%%%%%%%%%%%%%%%%%%%%%%%%%%%%%%%%%%%%%%%%% 
% \noindent{\bf Virtual Sensors}. 
Virtual sensors (e.g., \cite{Aberer2006,brunelli2016sensormind,gupta2016rationale,kim2017scalable,dautov2017three,dautov2019hierarchical}) are used to collect data from one or more physical sensors. A user-defined function can be applied to produce a measurement that applications can use by sending a query to a database where the sensor data is stored. Held \textit{et al.} \cite{held2022systematic} highlighted the following IoT characteristics relevant for middleware: 1) Scalability, 2) Heterogeneity, 3) self-configuration, 4) self-discovering, 5) self-processing, 5) Everything-as-a-Service, and 6) Security.

% \noindent{\bf IoT Platforms}. 
There is considerable work on IoT platforms (e.g., \cite{pintus2012paraimpu,sinha2015xively,persson2015calvin,soldatos2015openiot,abdelwahab2015cloud,kim2016developing,o2017node,li2017virtual,elsaleh2019iot,ogawa2019performance}) focused on using sensor abstractions to hide the complexity of connecting to an IoT device and enabling semantic interoperability, which allows for highly heterogeneous IoT devices and services to exchange information and use the information exchanged.
 
Some IoT platforms focus on enabling developers/administrators to specify policies related to data workflow or placement of services.  For example, Hao \textit{et al.} \cite{hao2017challenges} propose a software architecture, WM-FOG, that provides a language for developers to specify workflow policies that may make better use of the underlying hardware resources.  Giang \textit{et al.} \cite{giang2018exogenous} focus on specified criteria to be used in deployment, e.g., a constraint specifying a geographical area where a service can be deployed.  This work allows for services to be used in multiple applications but does not enable a service instance to be used by different applications. 

% Several IoT platforms (e.g., \cite{alarbi2018sensing,detti2019virtual,habenicht2022syncmesh})  have been developed to focus on decoupling IoT infrastructure providers from application developers by providing sensor data as a service with the goal of sharing data with different applications.
Several IoT platforms (e.g., \cite{alarbi2018sensing,detti2019virtual,habenicht2022syncmesh})  have been developed to focus on decoupling IoT infrastructure providers from application developers by providing sensor data as a service and sharing it with different applications.  

% Zhang \textit{et al.} \cite{zhang2020tinyedge} allow developers to select modules and specify configurations by using \textit{TinyEdge}. TinyEdge parses the code into different parts, generates message queue topics, and distributes each part to designated execution modules or engines. There is no discussion on the design of the services, and it seems to assume a limited set of already defined services primarily.
Zhang \textit{et al.} \cite{zhang2020tinyedge} allow developers to select modules and specify configurations by using \textit{TinyEdge}. TinyEdge parses the code into different parts, generates message queue topics, and distributes each part to designated execution modules or engines. There is no discussion on the design of the services, and it assumes a limited set of defined services.

There are IoT platforms used to support the dataflow-oriented application model. For example, Cheng \textit{et al.} \cite{cheng2017fogflow} present FogFlow, which supports a data flow-oriented application model based on external configurations that specify how the data flows between the fog nodes through the use of tasks. These tasks subscribe to a data source that can be either a sensor or a form of an analytics task. A broker receives the data, then transmits it to the task that has subscribed to it. This approach means that data transfers are through multiple hops, which increases latency. There is little support for sharing service instances among IoT applications. Ogawa \textit{et al.} \cite{ogawa2019iot} allow for raw sensor sharing through typical virtual sensors representing a physical sensor. Zhang \textit{et al.} \cite{zhang2018firework} present their Firework framework that facilities data sharing and processing by sending functions to the data. We consider this complementary to our work; however, it does not appear to support the composition of multiple service topologies.

% \noindent{\bf Placement}. 
% There is work with a focus on the placement of IoT applications’ services \cite{brogi2017qos,taneja2017resource} based on QoS metrics and the current resource usage of the fog nodes. This is complementary to our work because some methods can be used in a future version of our middleware for optimal placement. In contrast, our work focuses on the deployment after a decision has been made on the placement.
There is work with a focus on the placement of IoT applications' services \cite{brogi2017qos,taneja2017resource} based on QoS metrics and the current resource usage of the fog nodes, which is complementary to our work because some methods can be used in a future version of our middleware for optimal placement. In contrast, our work focuses on the deployment after a decision has been made on the placement.

%\noindent{\bf Ease of programming}
% Improving programming productivity is the focus of several IoT platforms productivity \cite{zhang2020tinyedge,jung2021oneos,coviello2022dataxe}. For example, Zhang \textit{et al.} \cite{zhang2020tinyedge} allow developers to select modules and specify configurations.  Jung \textit {et al} \cite{jung2021oneos} focus on expressing a sequence of services as a pipeline.  The work presents a high level language that runs on a Posix-compliant platform. Coviello et al \cite{coviello2022dataxe} shows that programming constructs improve programming productivity.  Some aspects of this work can be be used in our work to improve programming productivity.
Improving programming productivity is the focus of several IoT platforms' productivity \cite{zhang2020tinyedge,jung2021oneos,coviello2022dataxe}. For example, Zhang \textit{et al.} \cite{zhang2020tinyedge} allow developers to select modules and specify configurations. Jung \textit {et al.} \cite{jung2021oneos} focus on expressing a sequence of services as a pipeline.  The work presents a high-level language that runs on a Posix-compliant platform. Coviello \textit{et al.} \cite{coviello2022dataxe} show that programming constructs improve programming productivity. Some aspects of this work can be used in our work to improve programming productivity.

%Habenicht et al \cite{habenicht2022syncmesh} designed SynchMesh which provides a query language that assumes sensor data is placed on fog nodes and programmers can specify queries.

% \noindent{\bf Commercial Platforms}. 
Our review of commercial IoT platforms and frameworks (e.g., \cite{platform-awsiot2019,platform-azureiot2019,platform-googlecloud2019,platform-fiware2019}) finds that most of the platforms provide support for integrating devices into the IT structure with a diverse set of protocol connectors to deal with heterogeneous protocols. There is often a way to add new protocol connectors as needed \cite{platform-fiware2019}. Several platforms \cite{platform-awsiot2019,platform-azureiot2019,platform-googlecloud2019,platform-fiware2019} allow for processing of sensor data on an fog node but use the cloud for durable storage and analytics. For example, AWS IoT Greengrass \cite{platform-awsiot2019} can run AWS Lambda functions on sensor data on a fog node.

% \noindent \textbf{Contribution}. 
Overall the current work provides little support for sharing service instances among IoT applications and does not provide support for handling data from the perspective of dealing with the rate of receiving data from multiple sources and dealing with missing data.

%%%%%%%%%%%%%%%%%%%%%%%%%%%%%%%%%%%%%%%%%%%%%%%%%%%%%%%%%%%%%%%%%%%%%%%%%%%%%%%%%%%%%%%
%%%%%%%%%%%%%%%%%%%%%%%%%%%%%%%%%%%%%%%%%%%%%%%%%%%%%%%%%%%%%%%%%%%%%%%%%%%%%%%%%%%%%%%  
\section{Virtual Sensor Middleware Architecture} \label{sec:arch}
%%%%%%%%%%%%%%%%%%%%%%%%%%%%%%%%%%%%%%%%%%%%%%%%%%%%%%%%%%%%%%%%%%%%%%%%%%%%%%%%%%%%%%%
%%%%%%%%%%%%%%%%%%%%%%%%%%%%%%%%%%%%%%%%%%%%%%%%%%%%%%%%%%%%%%%%%%%%%%%%%%%%%%%%%%%%%%% 
This section describes the virtual sensor design and the middleware design as described in AlMahamid published thesis \cite{almahamid2019virtual}.
%=====================================================================================%
\subsection{Virtual Sensor}
%=====================================================================================% 
A Virtual Sensor (VS) provides a layer of abstraction to user applications such that application developers do not need to deal with communication technologies \cite{gupta2016implementation}. A VS is typically used as a software representation of a physical sensor that uses an adapter for connection management to receive the physical sensor's data. Different adapters can enable communication with a physical sensor over different protocols. An IoT application service may need to process data from one or more data streams and output a data stream. VSM uses virtual sensors to receive data from one or more data streams(i.e., aggregate data received from multiple sources) and output a data stream. The resulting output is based on the processing of input data.

Each virtual sensor is instantiated based on an input configuration representing attribute values stored in a configuration file. The configuration attribute values specify the information needed to operate, e.g., input sources, output destination, reading rate from input queues, fault handling policy, and database address. The rest of this section describes the communication paradigm used between virtual sensors and virtual sensor components.

%-------------------------------------------------------------------------------------%
\subsubsection{Communications Between Virtual Sensors}
%-------------------------------------------------------------------------------------%
Publish-Subscribe is a messaging pattern that provides a bidirectional messaging strategy in which subscribers may receive information asynchronously in the form of messages from publishers via a message broker without publishers having to be aware of subscribers and vice versa.

Publishers and subscribers are completely decoupled in time, space, and synchronization via Publish-Subscribe interaction \cite{happ2016limitations}.
A subscriber gets a subset of all messages via publish-subscribe. One method for defining the subset of communications is for the publisher to assign a subject to each message, and for subscribers to use the topic to filter messages \cite{alarbi2018sensing}. A virtual sensor that functions as both a publisher and a subscriber represents an IoT application service with both incoming and outgoing connections. A service with no incoming edge is assumed to receive data from a physical sensor directly. In contrast, a service with no outgoing edge cannot publish data. 

The publish-subscribe paradigm enables a virtual sensor to be developed without worrying about the location of other virtual sensors, which makes it easier to integrate virtual sensors into an application and allows a virtual sensor to be used by multiple applications.

%-------------------------------------------------------------------------------------%
\subsubsection{Virtual Sensor Components}\label{sec:vs-components}
%-------------------------------------------------------------------------------------%
The virtual sensor is instantiated based on the configuration file, which specifies how the virtual sensor operates. VS components do not need to be changed since the input sources used by the consumer are defined in the configuration file. Only the configuration file needs to be updated if a change is required to a VS, such as updating the input sources.

\begin{figure}[!t]
	\centering
	\includegraphics[width=1\linewidth]{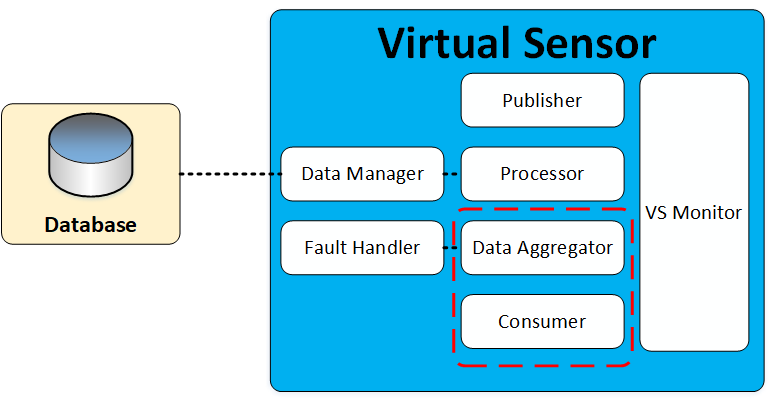}
	\caption{Virtual Sensor Components}
	\label{fig:vs-components}
% 	\vspace{-3mm}
\end{figure}

The virtual sensor consists of multiple components, as illustrated in Figure \ref{fig:vs-components}:\\
\noindent \textbf{Consumer}: establishes a connection with the input source and consumes the data received from the input source. In case of the input source is a virtual sensor, then the Consumer will subscribe to the message broker on the fog node that hosts the virtual sensor publishing the data. Nevertheless, if the input source is a physical sensor, adapters are employed to create connections with the physical sensor. The Consumer will start receiving data only after the connection is established.

\noindent \textbf{Data Aggregator}: combines data from various input sources, which might be received at different rates. The \textit{Data Aggregator} employs a mechanism that creates a priority queue per input source, then sorts messages/data depending on their timestamps in the created priority queues. The \textit{Data Aggregator} dequeues and saves all accessible data from priority queues in a tuple at a rate set by the virtual sensor configuration. If the created tuple holds incomplete data from the priority queues, then the \textit{Fault Handler} decides how to proceed in the event of an error. Nevertheless, if the tuple has no issues, then the \textit{Data Aggregator} transfers the data to the \textit{Processor} component.

\noindent \textbf{Fault Handler}: is activated by the \textit{Data Aggregator} when data are missing or are not in the proper format. The \textit{Fault Handler} verifies the fault handling policy, which determines the action on how to proceed with data processing. For instance, when data is missing, the application may continue processing incomplete data or wait for the missing data to come. 

\noindent \textbf{Processor}: carries out the desired functionality of the VS by processing the incoming tuple. For instance, it can apply aggregate functions such as the mean or sum. Once the \textit{Processor} finishes processing the data, it transmits the output to the \textit{Publisher}.

\noindent \textbf{Publisher}: is responsible for establishing a connection with the message broker in order to publish the data generated by the \textit{Processor} in a manner where subscribers may access it. 

\noindent \textbf{Data Manager}: connects to the database depending on the virtual sensor setup if data storage is required, which might be helpful if applications require access to past data generated by the virtual sensor.

\noindent \textbf{VS Monitor}: regularly transmits messages to the cloud-based VSM Monitor. The lack of a message is used by the \textit{VSM Monitor} to identify that the virtual sensor is offline. 

%=====================================================================================%
\subsection{Middleware Components}
\begin{figure*}[!t]
	\centering
	\includegraphics[width=1.0\linewidth]{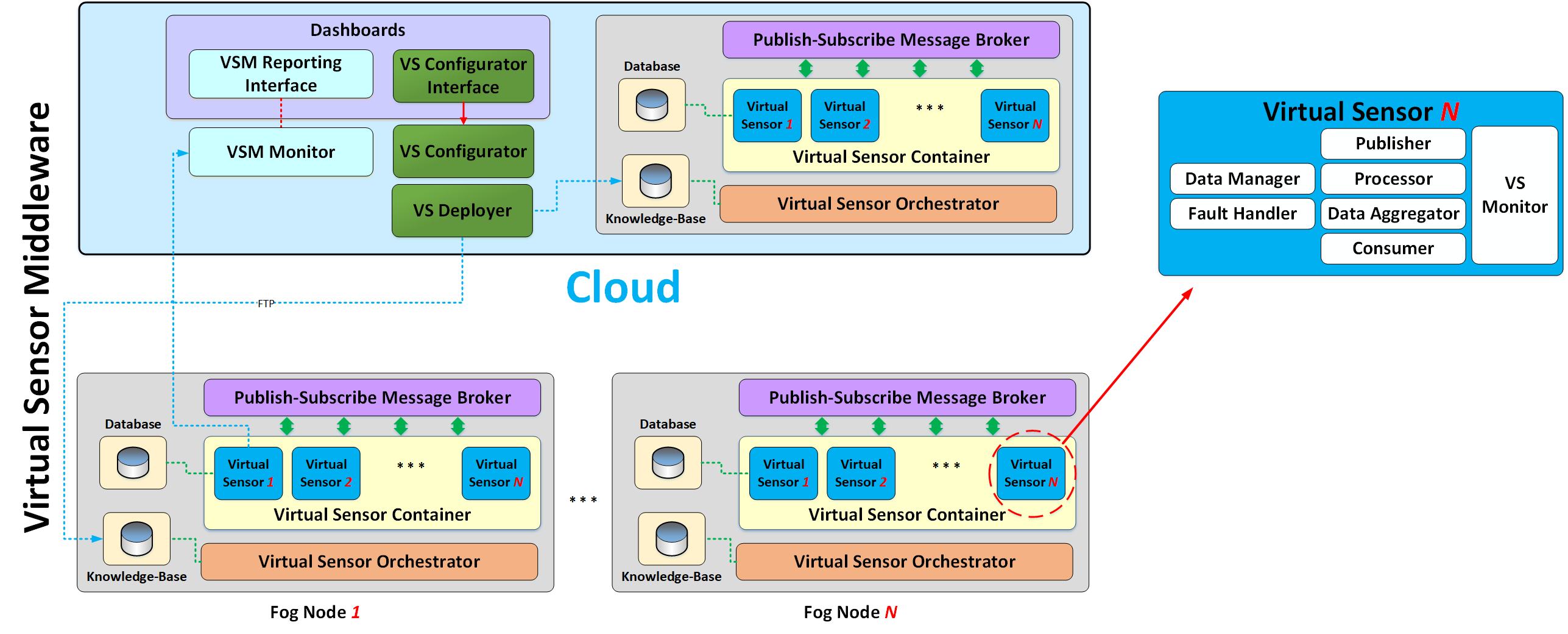}
	\caption{VS Middleware Architecture}
	\label{middleware-architecture}
% 	\vspace{-1mm}
    \vspace{-5mm}
\end{figure*}

%=====================================================================================%
This section describes the architectural components depicted in Figure \ref{middleware-architecture} associated with the cloud and fog nodes. 

%-------------------------------------------------------------------------------------%
\subsubsection{Middleware Fog Components}
%-------------------------------------------------------------------------------------%
Each fog node has a \textit{publish-subscribe message broker} that accepts data from the hosted virtual sensors. Regardless of where the virtual sensor is located, it can subscribe to the \textit{publish-subscribe broker} if interested in the published data. The \textit{knowledge-base} includes settings for all virtual sensors deployed and hosted on the fog node. These settings are used to instantiate virtual sensors using the \textit{VS Orchestrator}, which collects information from the \textit{knowledge-base} regarding newly added or updated virtual sensor configurations. The \textit{VS Orchestrator} instantiates or updates the virtual sensor depending on the \textit{knowledge-base} configuration and keeps a list of all instantiated virtual sensors. 

%-------------------------------------------------------------------------------------%
\subsubsection{Middleware Cloud Components}
%-------------------------------------------------------------------------------------%
A script or graphical user interface (GUI) is utilized to set up the virtual sensor configurations and select the hosting nodes. The cloud hosts the VS Configurator, which obtains configuration data through the VS Configurator Interface. It generates and verifies the syntax of the virtual sensor setups and stores them in a file.\\
\noindent \textbf{VS Deployer} is responsible for distributing the configuration file to the specified fog nodes following the creation of the configuration file.\\ 
\noindent \textbf{VSM Monitor} is responsible for monitoring the status of all deployed virtual sensors, which are exclusively utilized for reporting. Each virtual sensor reports its status to the VSM Monitor at the set interval. The VSM Monitor considers that the virtual sensor is offline if it does not receive a message from it.

%%%%%%%%%%%%%%%%%%%%%%%%%%%%%%%%%%%%%%%%%%%%%%%%%%%%%%%%%%%%%%%%%%%%%%%%%%%%%%%%%%%%%%%
%%%%%%%%%%%%%%%%%%%%%%%%%%%%%%%%%%%%%%%%%%%%%%%%%%%%%%%%%%%%%%%%%%%%%%%%%%%%%%%%%%%%%%% 
\section{Middleware Interactions} \label{sec:interactions}  
%%%%%%%%%%%%%%%%%%%%%%%%%%%%%%%%%%%%%%%%%%%%%%%%%%%%%%%%%%%%%%%%%%%%%%%%%%%%%%%%%%%%%%%
%%%%%%%%%%%%%%%%%%%%%%%%%%%%%%%%%%%%%%%%%%%%%%%%%%%%%%%%%%%%%%%%%%%%%%%%%%%%%%%%%%%%%%% 

This section describes the interactions of the components through scenarios.

%=====================================================================================%
\subsection{Creating and Instantiating Virtual Sensor}
%=====================================================================================%
Virtual sensor configurations are deployed to a specified node, i.e., fog nodes or the cloud. The configurations are used to instantiate the virtual sensor. VS configurations is specified using \textit{VS Configurator Interface}, which forwarded to \textit{VS Configurator} to generate a configuration file. The configuration file will be transferred to the \textit{Knowledge-Base} at the destination node using \textit{VS Deployer}. \textit {VS Orchestrator} at the destination node will be watching the \textit{Knowledge-Base} for updates, i.e., a new file is added, or an existing file is updated. The \textit{VS Orchestrator} instantiates the virtual sensor using the configuration file. Once the virtual sensor is instantiated, it can start exchanging and processing data.

%=====================================================================================%
\subsection{Exchanging Messages Between Virtual Sensors}
%=====================================================================================%
We use a  scenario to describe the interaction between two virtual sensors. The first virtual sensor, VS1, publishes data, while the second virtual sensor, VS2, receives the published data after subscribing to the message broker. The scenario assumes that the data is ready for publishing at VS1, and no fault-handling is required and ends when the VS2 publishes the consumed data from VS1.

It starts when \textit{VS1 Publisher} timestamps the message and publishes it to the VS1 topic declared at \textit{Message Broker}, then the \textit{Message Broker}  notifies all subscribers and forwards the published message. \textit{VS2 Consumer} receives the published message since it is registered as a subscriber. The \textit{VS2 Consumer} timestamps a message and adds it to a designated priority-queue that is used to store data received from VS1.VS2 has a designated queue per input source. The \textit{VS2 Data Aggregator} runs at a configured rate to check for data in the queues. Once it reads the data from the queues, it creates a tuple with the data found from all queues and forwards the tuple to the \textit{Processor}. Upon receiving the tuple, the \textit{Processor} applies the aggregation function or filters per the VS2 configurations. It then forwards the result to the \textit{VS2 Publisher} where it timestamps the data and publishes it to the \textit{Message Broker}. VS1 and VS2 can be located on the same or different fog nodes.

%%%%%%%%%%%%%%%%%%%%%%%%%%%%%%%%%%%%%%%%%%%%%%%%%%%%%%%%%%%%%%%%%%%%%%%%%%%%%%%%%%%%%%%
%%%%%%%%%%%%%%%%%%%%%%%%%%%%%%%%%%%%%%%%%%%%%%%%%%%%%%%%%%%%%%%%%%%%%%%%%%%%%%%%%%%%%%%  
\section{Implementation} \label{sec:Implementation}
%%%%%%%%%%%%%%%%%%%%%%%%%%%%%%%%%%%%%%%%%%%%%%%%%%%%%%%%%%%%%%%%%%%%%%%%%%%%%%%%%%%%%%%
%%%%%%%%%%%%%%%%%%%%%%%%%%%%%%%%%%%%%%%%%%%%%%%%%%%%%%%%%%%%%%%%%%%%%%%%%%%%%%%%%%%%%%%
The implementation as described by AlMahamid \cite{almahamid2019virtual}: The \textit{Virtual Sensor Configurator Interface} is implemented using  Java Server Pages (JSP), JSON editor, and the Bootstrap framework.
The \textit{Virtual Sensor Configurator} component is a Java Servlet. A configuration is stored in a JSON file.  
The \textit{Virtual Sensor Deployer} component is a Java Servlet that uses Apache Commons Net v3.6 \cite{apachecommonnet} to transfer virtual sensor configurations (JSON file) to the desired node using File Transfer Protocol (FTP).

The Publish-Subscribe Message Broker is implemented using RabbitMQ, which is an open-source message broker that uses AMQP 0-9-1 as the underlying protocol to exchange messages between publishers and subscribers asynchronously e.g. \cite{rabbitmq2019amqp0-9-1}. In the current implementation, RabbitMQ is deployed on each fog node using the federation topology, where virtual sensors are deployed at different nodes and able to exchange messages using the RabbitMQ-federation. 

The \textit{Knowledge-Base} uses the filesystem to store all virtual sensor configurations (JSON files). Once a new file is added to the Knowledge-Base, it produces an event that is picked up by the \textit{VS Orchestrater} which is implemented in Java. The \textit{Virtual Sensor Orchestrater} instantiates the virtual sensor and adds it to the \textit{VS Container}. The \textit{VS Container} maintains all references of the instantiated Virtual Sensor objects. These references are used to update virtual sensors based on the virtual sensor configuration file updates.

%%%%%%%%%%%%%%%%%%%%%%%%%%%%%%%%%%%%%%%%%%%%%%%%%%%%%%%%%%%%%%%%%%%%%%%%%%%%%%%%%%%%%%%
%%%%%%%%%%%%%%%%%%%%%%%%%%%%%%%%%%%%%%%%%%%%%%%%%%%%%%%%%%%%%%%%%%%%%%%%%%%%%%%%%%%%%%% 
\section{Evaluation and Experiments Setup}\label{sec:Evaluation}
%%%%%%%%%%%%%%%%%%%%%%%%%%%%%%%%%%%%%%%%%%%%%%%%%%%%%%%%%%%%%%%%%%%%%%%%%%%%%%%%%%%%%%%
%%%%%%%%%%%%%%%%%%%%%%%%%%%%%%%%%%%%%%%%%%%%%%%%%%%%%%%%%%%%%%%%%%%%%%%%%%%%%%%%%%%%%%% 
The following subsections describes the experiments setup and the evaluation results.
%=====================================================================================%
\subsection{Experiments Setup}
\begin{figure}[!t]
	\centering
	\includegraphics[width=0.6\linewidth]{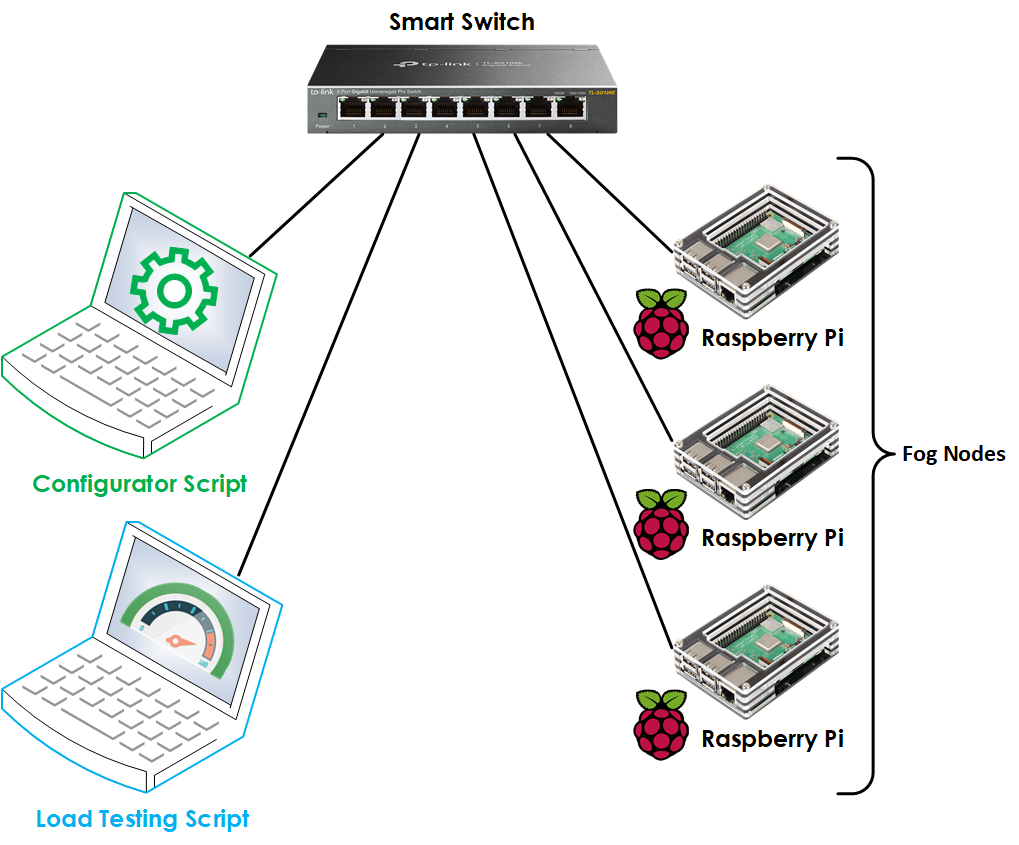}
	\caption{Evaluation Environment using Raspberry Pi serving as fog nodes and other scripts used for execute the simulation}
	\label{fig:evaluation-environment}
	\vspace{-5mm}
\end{figure}

\begin{figure}[!t]
	\centering
	\includegraphics[width=0.6\linewidth]{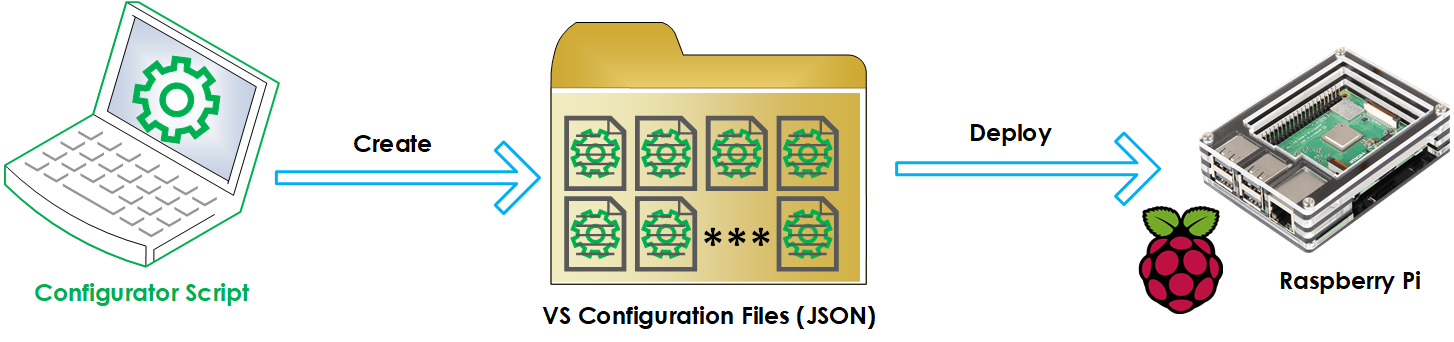}
	\caption{Evaluation Environment - Creating and deploying VS Configurations}
	\label{fig:evaluation-creating-configurations}
\end{figure}

\begin{figure}[!t]
	\centering
	\includegraphics[width=0.6\linewidth]{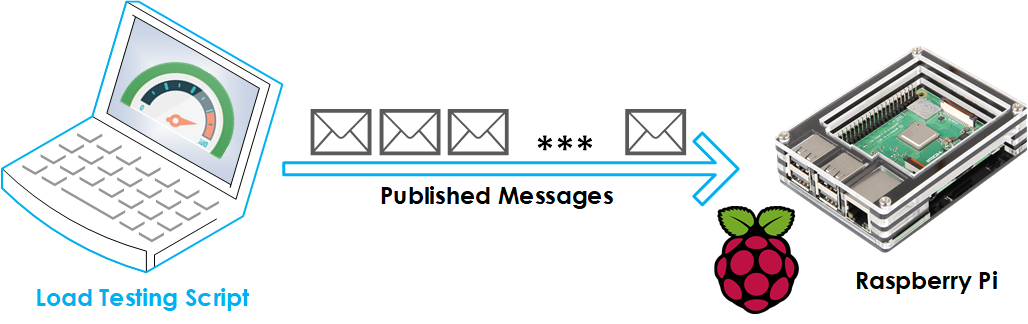}
	\caption{Evaluation Environment - Load Testing Script simulating data generated by physical sensors}
	\label{fig:evaluation-load-testing}
	\vspace{-5mm}
\end{figure}

%=====================================================================================%
As described in Section \ref{sec:arch}, the middleware consists of different components that run on the cloud or on the fog node. The cloud components provide the interface required to generate and deploy virtual sensors configuration where the fog nodes contain the different components responsible for processing sensor data. Therefore, the following are used to simulate the performance as shown in figure \ref{fig:evaluation-environment}:

\noindent \textbf{Raspberry Pi} is chosen to represent the fog nodes due to the low cost of such device and its modest processing capability. If the proof-of-concept is conducted on such a device, it can be scaled up to more powerful processing devices.\\
\noindent \textbf{Configuration Script} is used to construct a large number of virtual sensor configurations, describe their relationships, and deploy them to the Knowledge-Base at the selected fog node.
as shown in Figure \ref{fig:evaluation-creating-configurations}.\\  
\noindent \textbf{Load Testing Script} is used to simulate sensor data produced by physical sensors and transmit the data to the message broker of each participating fog node, as shown in Figure \ref{fig:evaluation-load-testing}.

%=====================================================================================%
\subsection{Evaluation}
%=====================================================================================%
The evaluation examines VSM behavior as the number of virtual sensors, the number of inputs per virtual sensor, the number of deployment levels used, and the number of fog nodes increases. The test environment uses a Raspberry PI to represent a fog node and RabbitMQ as a message broker. The evaluation is not meant to evaluate the message broker performance, i.e., RabbitMQ, and compare it to other message broker software, but it focuses on evaluating the middleware's performance by creating different virtual sensor setups. The virtual sensors are created, and their relations are defined in different setups and then deployed to the corresponding fog nodes.

%-------------------------------------------------------------------------------------%
\subsubsection{Number of Virtual Sensors}
\begin{figure}[!t]
    \centering
    \subfloat[CPU Comparison]{
    	\includegraphics[width=0.95\linewidth]{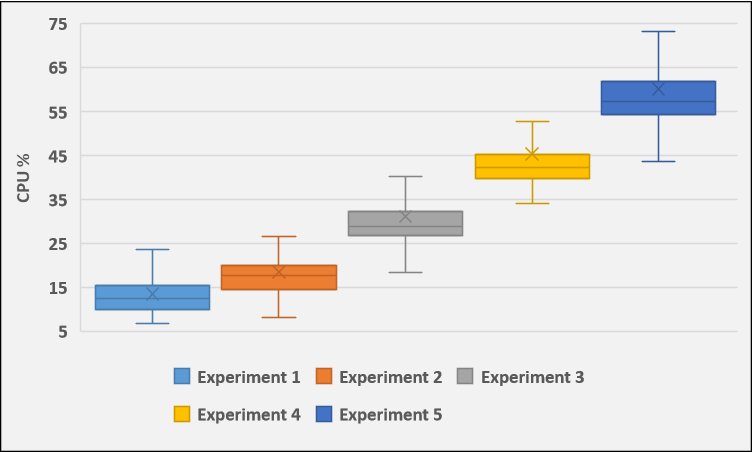}
    	\label{fig:scenario03_cpu_performance_comparison}
    }
	\vspace{2mm}
    \subfloat[Memory Comparison]{
    	\includegraphics[width=0.95\linewidth]{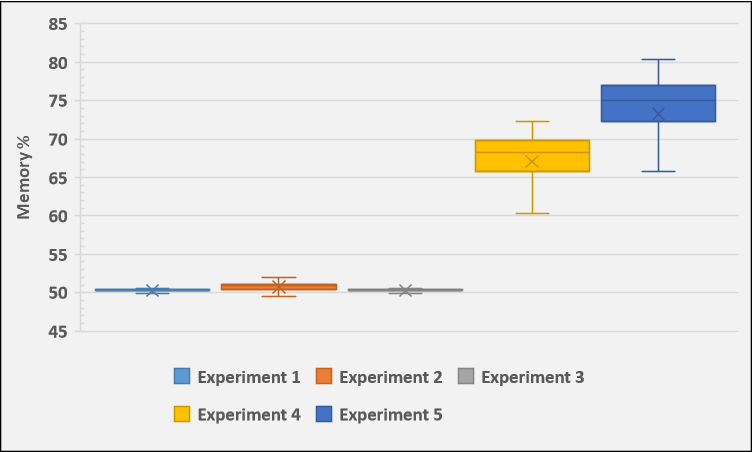}%
    	\label{fig:scenario03_memory_performance_comparison}
    }
    \caption{Number of Virtual Sensor Impact on Performance}
    \label{fig:scenario03_comparison}
    \vspace{-5mm}
\end{figure}

%-------------------------------------------------------------------------------------%
The number of virtual sensors used can vary depending on the requirements. Therefore, this experiment aims to show the performance behavior on a single node when the number of virtual sensors increases. Each virtual sensor was configured to receive a single input from a given source. Table \ref{tab:vs-cpu} \& \ref{tab:vs-mem}, and Figure \ref{fig:scenario03_comparison} show that CPU and memory utilization increased when the number of virtual sensors increased. This behavior is expected since each virtual sensor requires resources to process sensor data.

\begin{table}[!t]
	\centering
	\caption{Impact of number of virtual sensors on the CPU}
	\label{tab:vs-cpu}
	\begin{tabular}{lccccc}
		\toprule
		No. of VSs&Min&Max&Average&Median&SD.\\
		\midrule
		1&$6.78\%$&$57.03\%$&$13.51\%$&	$12.41\%$&$6.02\%$\\
		10&$8.14\%$&$85.90\%$&$18.50\%$&$17.77\%$&$8.82\%$\\
		50&$17.75\%$&$94.67\%$&$31.14\%$&$28.93\%$&$10.75\%$\\
		100&$16.47\%$&$95.90\%$&$45.29\%$&$45.26\%$&$11.35\%$\\
		150&$18.61\%$&$98.22\%$&$60.18\%$&$57.25\%$&$10.45\%$\\
		\bottomrule
	\end{tabular}
    \vspace{2mm}
	\caption{Impact of number of virtual sensors on the Memory}
	\label{tab:vs-mem}
	\begin{tabular}{lccccc}
		\toprule
		No. of VSs&Min&Max&Average&Median&SD.\\
		\midrule
		1&$46.27\%$&$51.73\%$&$50.28\%$&$50.31\%$&$0.39\%$\\
		10&$45.71\%$&$52.65\%$&$50.71\%$&$50.91\%$&$0.63\%$\\
		50&$44.57\%$&$59.14\%$&$56.14\%$&$56.43\%$&$1.90\%$\\
		100&$45.05\%$&$72.26\%$&$67.08\%$&$68.27\%$&$4.39\%$\\
		150&$45.38\%$&$80.40\%$&$73.25\%$&$75.08\%$&$6.29\%$\\
		\bottomrule
	\end{tabular}
\end{table}

%-------------------------------------------------------------------------------------%
\subsubsection{Total Number of Input Sources per Virtual Sensor}
\begin{figure}[!t]
\centering
	\subfloat[CPU Comparison]{
		\includegraphics[width=0.95\linewidth]{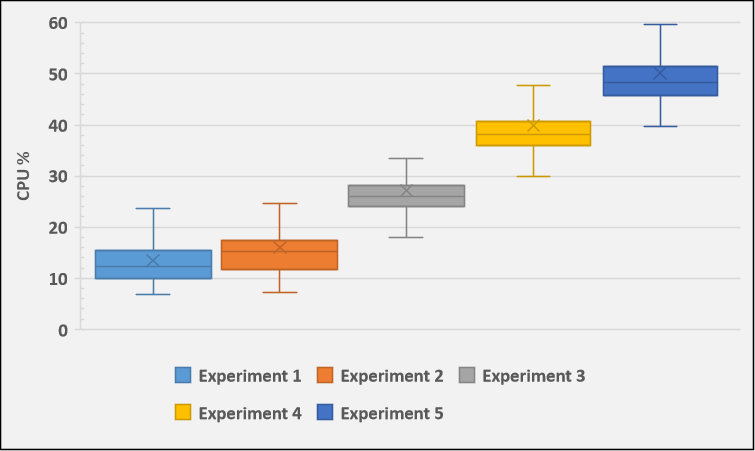}
		\label{fig:scenario02_cpu_performance_comparison}
	}
	\vspace{2mm}
	\subfloat[Memory Comparison]{
		\includegraphics[width=0.95\linewidth]{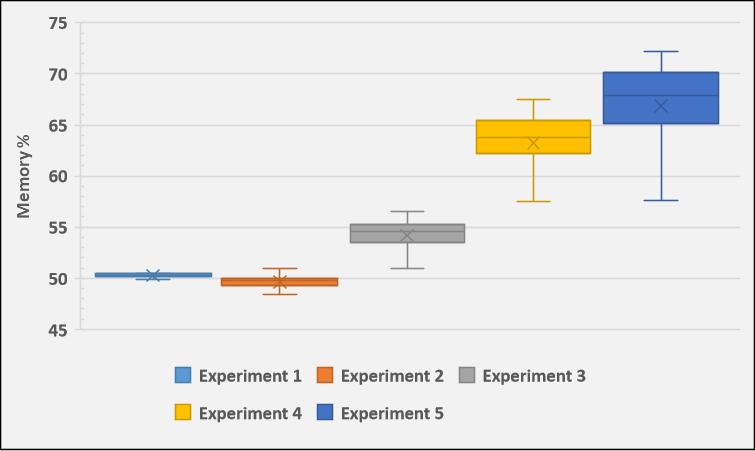}
		\label{fig:scenario02_memory_performance_comparison}
	}
	\caption{Number of Inputs per Virtual Sensor Impact on Performance}
	\label{fig:scenario02_comparison}
	\vspace{-5mm}
\end{figure}

%-------------------------------------------------------------------------------------%
The virtual sensor can be configured to receive inputs from different physical and virtual sensors. Therefore, to illustrate the impact on the performance, a single virtual sensor is tested with different input sources, i.e., 1, 10, 50, 100, and 150. Tables \ref{tab:input-sources-cpu} \& \ref{tab:input-sources-mem}, and Figure \ref{fig:scenario02_comparison} show that the CPU and memory utilization increased as the number of input sources increases. The utilization increase is justified since the message broker declares an independent topic for each input source. Furthermore, the virtual sensor needs to define a queue to store received data for each input source. We can infer that the more input sources a virtual sensor has, the more CPU and memory are required.

\begin{table}[!ht]
	\centering
	\caption{Impact of number of inputs per virtual sensor on the CPU}
	\label{tab:input-sources-cpu}
	\begin{tabular}{lccccc}
		\toprule
		No. of Inputs&Min&Max&Average&Median&SD.\\
		\midrule
		1&$6.78\%$&$57.03\%$&$13.51\%$&$12.41\%$&$6.02\%$\\
		10&$7.20\%$&$66.84\%$&$16.01\%$&$15.19\%$&$0.22\%$\\
		50&$15.80\%$&$65.64\%$&$27.22\%$&$26.02\%$&$7.10\%$\\
		100&$15.57\%$&$82.52\%$&$39.91\%$&$38.17\%$&$7.79\%$\\
		150&$21.28\%$&$85.90\%$&$50.16\%$&$48.30\%$&$7.71\%$\\
		\bottomrule
	\end{tabular}
    \vspace{2mm}
	\caption{Impact of number of inputs per virtual sensor on the Memory}
	\label{tab:input-sources-mem}
	\begin{tabular}{lccccc}
		\toprule
		No. of Inputs&Min&Max&Average&Median&SD.\\
		\midrule
		1&$46.27\%$&$51.73\%$&$50.28\%$&$50.31\%$&$0.39\%$\\
		10&$44.47\%$&$51.69\%$&$49.65\%$&$49.77\%$&$0.60\%$\\
		50&$44.35\%$&$56.54\%$&$54.21\%$&$54.60\%$&	$1.58\%$\\
		100&$48.34\%$&$67.46\%$&$63.25\%$&$63.82\%$&$2.99\%$\\
		150&$44.91\%$&$72.22\%$&$66.85\%$&$67.87\%$&$4.81\%$\\
		\bottomrule
	\end{tabular}
\end{table}

%-------------------------------------------------------------------------------------%
\subsubsection{Virtual Sensor Distribution on Multiple Levels of Fog Nodes}
\begin{figure}[!ht]
    \centering
	\subfloat[CPU Comparison]{
		\includegraphics[width=0.95\linewidth]{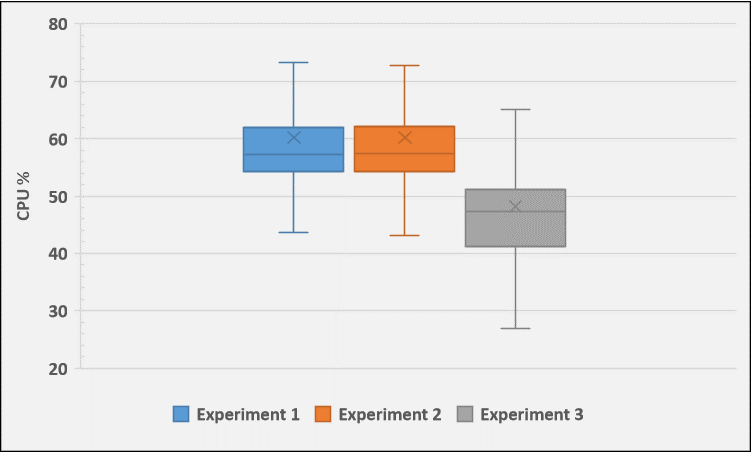}
		\label{fig:scenario04_cpu_performance_comparison}
	}
	\vspace{2mm}
	\subfloat[Memory Comparison]{
		\includegraphics[width=0.95\linewidth]{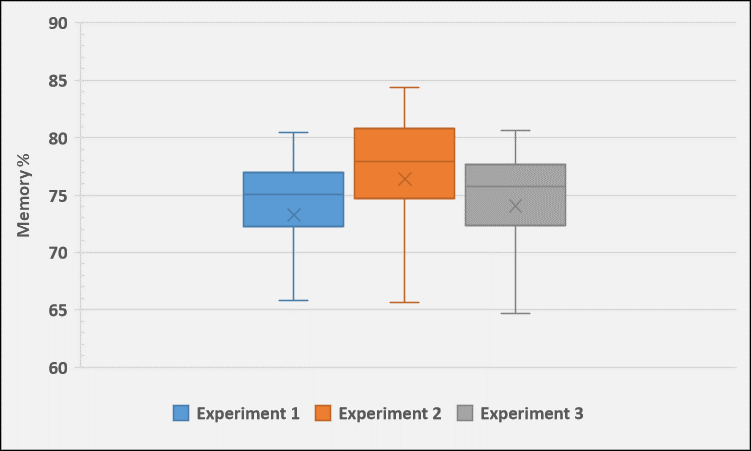}
		\label{fig:scenario04_memory_performance_comparison}
	}
	\caption{Deployment levels Impact on Performance}
	\label{fig:scenario04_comparison}
	\vspace{-5mm}
\end{figure}

%-------------------------------------------------------------------------------------%
Virtual sensors can be structured to process sensor cyclic-graph or in a structured tree, where each level passes the processed sensor data to the next upper level. This experiment aims to show the impact on the performance of the distribution of virtual sensors among multiple levels of a structured tree. One hundred and fifty virtual sensors are used in three different setups. The first setup assumes all virtual sensors are deployed at a single level. The second setup assumes that 100 virtual sensors are deployed at the first level, and 50 virtual sensors receive their inputs from the first level. The third setup distributes 150 virtual sensors equally into three levels. 
As illustrated in tables \ref{tab:levels-cpu} \& \ref{tab:levels-mem}, and Figure \ref{fig:scenario04_comparison} we observed that in the second setup, when we distributed the virtual sensors across two levels, there was a slight increase in the CPU and memory utilization. The utilization increase can be due to the increase in the number of input sources at level 2 since each virtual sensor requires to receive data from two sensors from the first level. However, when we distributed the virtual sensors equally into three levels in the third setup, we noticed that it decreased the CPU utilization, but the memory utilization almost remained the same compared to the first setup.

\begin{table}[!t]
	\centering

	\caption{Impact of deployment levels on the CPU}
	\label{tab:levels-cpu}
	\begin{tabular}{lccccc}
		\toprule
		No. of Levels&Min&Max&Average&Median&SD.\\
		\midrule
		1&$18.61\%$&$98.22\%$&$60.18\%$&$57.25\%$&$10.45\%$\\
		2&$15.09\%$&$99.25\%$&$60.20\%$&$57.44\%$&$10.90\%$\\
		3&$16.43\%$&$98.00\%$&$47.33\%$&$47.33\%$&$13.58\%$\\
		\bottomrule
	\end{tabular}
	\vspace{2mm}
	\caption{Impact of deployment levels on the Memory}
	\label{tab:levels-mem}
	\begin{tabular}{lccccc}
		\toprule
		No. of Levels&Min&Max&Average&Median&SD.\\
		\midrule
		1&$45.38\%$&$80.40\%$&$73.25\%$&$75.08\%$&$6.29\%$\\
		2&$44.98\%$&$84.37\%$&$76.38\%$&$77.95\%$&$7.30\%$\\
		3&$48.12\%$&$80.58\%$&$74.02\%$&$75.75\%$&$5.83\%$\\
		\bottomrule
	\end{tabular}
\end{table}

%-------------------------------------------------------------------------------------%
\subsubsection{Distributed Processing of Virtual Sensors Across Multiple Nodes}
%-------------------------------------------------------------------------------------%
This experiment examined the behavior of VSM in distributing the processing of virtual sensors. The first experiment has 150 virtual sensors deployed in a single fog node at a single level. In the second experiment, there were 100 virtual sensors hosted on fog 1 and 50 virtual sensors hosted on fog 2 that received inputs from the 100 virtual sensors at fog 1, which means each virtual sensor received two inputs. In the third experiment, there were 50 virtual sensors hosted on fog 1, 50 virtual sensors hosted on fog 2, and 50 virtual sensors hosted on fog 3. The virtual sensors at fog 2 receive data from the virtual sensors at fog 1, and the virtual sensors at fog 3 receive data from virtual sensors at fog 2. Tables \ref{tab:multi-fog-1} through \ref{tab:multi-fog-3} show the results of the three different experiments. We observed that CPU and memory utilization are reduced when we distribute the processing on other nodes.

\begin{table}[!t]
	\caption{CPU and Memory performance using one fog node}
	\label{tab:multi-fog-1}
	\begin{tabular}{lccccc}
		\toprule
		&Min&Max&Average&Median&SD.\\
		\midrule
		CPU Fog 1&$15.45\%$&$96.44\%$&$57.81\%$&$54.77\%$&$10.93\%$\\
		Memory Fog 1&$44.12\%$&$79.44\%$&$73.08\%$&	$74.08\%$&	$6.38\%$\\
		\bottomrule
	\end{tabular}
	\vspace{1mm}
	\caption{CPU and Memory  performance using two fog nodes}
	\label{tab:multi-fog-2}
	\begin{tabular}{lccccc}
		\toprule
		&Min&Max&Average&Median&SD.\\
		\midrule
		CPU Fog 1&$11.93\%$&$96.70\%$&$50.67\%$&$48.96\%$&$11.42\%$\\
		Memory Fog 1&$44.95\%$&	$73.72\%$&$67.86\%$&$69.16\%$&$4.85\%$\\
		\midrule
		CPU Fog 2&$7.32\%$&$87.88\%$&$18.16\%$&$17.13\%$&$7.05\%$\\
		Memory Fog 2&$43.44\%$&	$59.64\%$&$56.47\%$&$56.70\%$&$1.75\%$\\
		\bottomrule
	\end{tabular}
	\vspace{1mm}
	\caption{CPU and Memory  performance using three fog nodes}
	\label{tab:multi-fog-3}
	\begin{tabular}{lccccc}
		\toprule
		&Min&Max&Average&Median&SD.\\
		\midrule
		CPU Fog 1&$9.65\%$&	$92.98\%$&$31.78\%$&$29.06\%$&$11.93\%$\\
		Memory Fog 1&$45.19\%$&	$63.17\%$&$59.26\%$&$59.74\%$&$2.49\%$\\
		\midrule
		CPU Fog 2&$6.13\%$&	$91.88\%$&$16.18\%$&$14.68\%$&$0.44\%$\\
		Memory Fog 2&$44.70\%$&	$59.11\%$&$56.76\%$&$57.27\%$&$1.70\%$\\
		\midrule
		CPU Fog 3&$4.57\%$&	$89.06\%$&$12.43\%$&$11.17\%$&$8.35\%$\\
		Memory Fog 3&$46.27\%$&$57.63\%$&$55.54\%$&$55.67\%$&$1.10\%$\\
		\bottomrule
	\end{tabular}
	\vspace{-2mm}
\end{table}

%%%%%%%%%%%%%%%%%%%%%%%%%%%%%%%%%%%%%%%%%%%%%%%%%%%%%%%%%%%%%%%%%%%%%%%%%%%%%%%%%%%%%%%
%%%%%%%%%%%%%%%%%%%%%%%%%%%%%%%%%%%%%%%%%%%%%%%%%%%%%%%%%%%%%%%%%%%%%%%%%%%%%%%%%%%%%%%  
\section{Conclusions and Future Work}\label{sec:Conclusion}
%%%%%%%%%%%%%%%%%%%%%%%%%%%%%%%%%%%%%%%%%%%%%%%%%%%%%%%%%%%%%%%%%%%%%%%%%%%%%%%%%%%%%%%
%%%%%%%%%%%%%%%%%%%%%%%%%%%%%%%%%%%%%%%%%%%%%%%%%%%%%%%%%%%%%%%%%%%%%%%%%%%%%%%%%%%%%%%
Important features include the following:
(i) The middleware design includes components for robustness purposes, which includes using a fault-handling policy to deal with the absence of sensor data that a  physical sensor may cause, virtual sensor a fog node going offline. Requiring that a virtual sensor periodically sends a message to a cloud component allows for a timely response to failures. If the \textit{VSM Monitor} (cloud component) does not a receive signals from all the virtual sensors (through the \textit{VS Monitor}) on a single fog node, then the \textit{VSM Monitor} could assume that the fog node is offline;  (ii) The middleware eliminates the dependency between IoT devices and applications through the use of the publish-subscribe design pattern, where data producers do not need to maintain information about data consumers. (iii) IoT applications can change at run-time. For example, a virtual sensor may need a new input stream. A new configuration is generated and sent to the knowledge-base. Changes can be identified, and the \textit{VS-Orchestrator} responds to notifications of changes by either subscribing or unsubscribing to the appropriate broker.  
Future work includes the following: (i) Currently, our work assumes that the \textit{Processor} code includes all the functions that can be used, e.g., filtering, aggregation, and average. We want to develop a more flexible approach where the virtual sensor uses an attribute that specifies the software to be downloaded and used by the virtual sensor; (ii) This work assumed that each fog node had a publish-subscribe broker. We want to compare different approaches to placing brokers, e.g., a fog node representing a region may have a publish-subscriber broker while others may not.

%%%%%%%%%%%%%%%%%%%%%%%%%%%%%%%%%%%%%%%%%%%%%%%%%%%%%%%%%%%%%%%%%%%%%%%%%%%%%%%%%%%%%%%
%%%%%%%%%%%%%%%%%%%%%%%%%%%%%%%%%%%%%%%%%%%%%%%%%%%%%%%%%%%%%%%%%%%%%%%%%%%%%%%%%%%%%%%  
\section*{Acknowledgment}
%%%%%%%%%%%%%%%%%%%%%%%%%%%%%%%%%%%%%%%%%%%%%%%%%%%%%%%%%%%%%%%%%%%%%%%%%%%%%%%%%%%%%%%
%%%%%%%%%%%%%%%%%%%%%%%%%%%%%%%%%%%%%%%%%%%%%%%%%%%%%%%%%%%%%%%%%%%%%%%%%%%%%%%%%%%%%%%
This research has been supported by NSERC under grants RPGIN-2017-02461 \& RGPIN-2018-06222.

%%%%%%%%%%%%%%%%%%%%%%%%%%%%%%%%%%%%%%%%%%%%%%%%%%%%%%%%%%%%%%%%%%%%%%%%%%%%%%%%%%%%%%%
%%%%%%%%%%%%%%%%%%%%%%%%%%%%%%%%%%%%% References %%%%%%%%%%%%%%%%%%%%%%%%%%%%%%%%%%%%%%
%%%%%%%%%%%%%%%%%%%%%%%%%%%%%%%%%%%%%%%%%%%%%%%%%%%%%%%%%%%%%%%%%%%%%%%%%%%%%%%%%%%%%%% 
\bibliographystyle{IEEEtran}
\bibliography{references}
%%%%%%%%%%%%%%%%%%%%%%%%%%%%%%%%%%%%%%%%%%%%%%%%%%%%%%%%%%%%%%%%%%%%%%%%%%%%%%%%%%%%%%%
%%%%%%%%%%%%%%%%%%%%%%%%%%%%%%%%%%%%%%%%%%%%%%%%%%%%%%%%%%%%%%%%%%%%%%%%%%%%%%%%%%%%%%%

\end{document}